\documentclass[runningheads]{llncs}
\usepackage[T1]{fontenc}
\usepackage{amsmath}
\usepackage{hyperref}
\usepackage{amsfonts}
\usepackage{graphicx}
\usepackage{booktabs}
\usepackage{array}
\usepackage{makecell}
\usepackage{booktabs}
\usepackage{float}
\usepackage{orcidlink}
\usepackage{marvosym}
\usepackage{makecell}
\usepackage{arydshln}

\usepackage{graphicx}
\begin{document}
\title{The Missing Evaluation Axis: What 10,000 Student Submissions Reveal About AI Tutor Effectiveness}
\authorrunning{R. Niousha et al.}
\titlerunning{The Missing Evaluation Axis}
%
%

\author{Rose Niousha$^{(\textrm{\Letter})}$\inst{1}\orcidlink{0000-0003-2587-6079} \and
Samantha Boatright Smith\inst{1}\orcidlink{0009-0004-8182-2424} \and
Bita Akram\inst{2}\orcidlink{0000-0001-5195-5841} \and
Peter Brusilovsky\inst{3}\orcidlink{0000-0002-1902-1464} \and
Arto Hellas\inst{4}\orcidlink{0000-0001-6502-209X} \and
Juho Leinonen\inst{4}\orcidlink{0000-0001-6829-9449} \and
John DeNero\inst{1}\orcidlink{0000-0001-9152-3891} \and
Narges Norouzi\inst{1}\orcidlink{0000-0001-9861-7540}}
%
%
\institute{University of California, Berkeley, USA
\email{\{rose.n,happysammie6,denero,norouzi\}@berkeley.edu} \and
North Carolina State University, USA\\
\email{bakram@ncsu.edu} \and
University of Pittsburgh, USA\\
\email{peterb@pitt.edu} \and
Aalto University, Finland\\
\email{\{arto.hellas,juho.2.leinonen\}@aalto.fi}}

\maketitle              
\begin{abstract}
Current Artificial Intelligence (AI)-based tutoring systems (AI tutors) are primarily evaluated based on the pedagogical quality of their feedback messages. While important, pedagogy alone is insufficient because it ignores a critical question: \textbf{what do students actually do with the feedback they receive?} We argue that AI tutor evaluation should be extended with a behavioral dimension grounded in student interaction data, which complements pedagogical assessment. We propose an evaluation framework and apply it to 10,235 code submissions with corresponding AI tutor feedback from an introductory undergraduate programming course to measure whether students act on tutor feedback and whether those actions are applied correctly. Using this framework to compare two deployed AI tutors across different semesters in a large-scale introductory computer science course reveals substantial differences in student engagement patterns that are not captured by pedagogy-only evaluation. Moreover, these engagement-based behavioral signals are more strongly associated with student perception of helpful feedback than pedagogical quality alone, providing a more complete and actionable picture of AI tutor performance.

\end{abstract}
\section{Introduction}
\begin{figure}[t]
    \centering
    \includegraphics[width=0.85\linewidth]{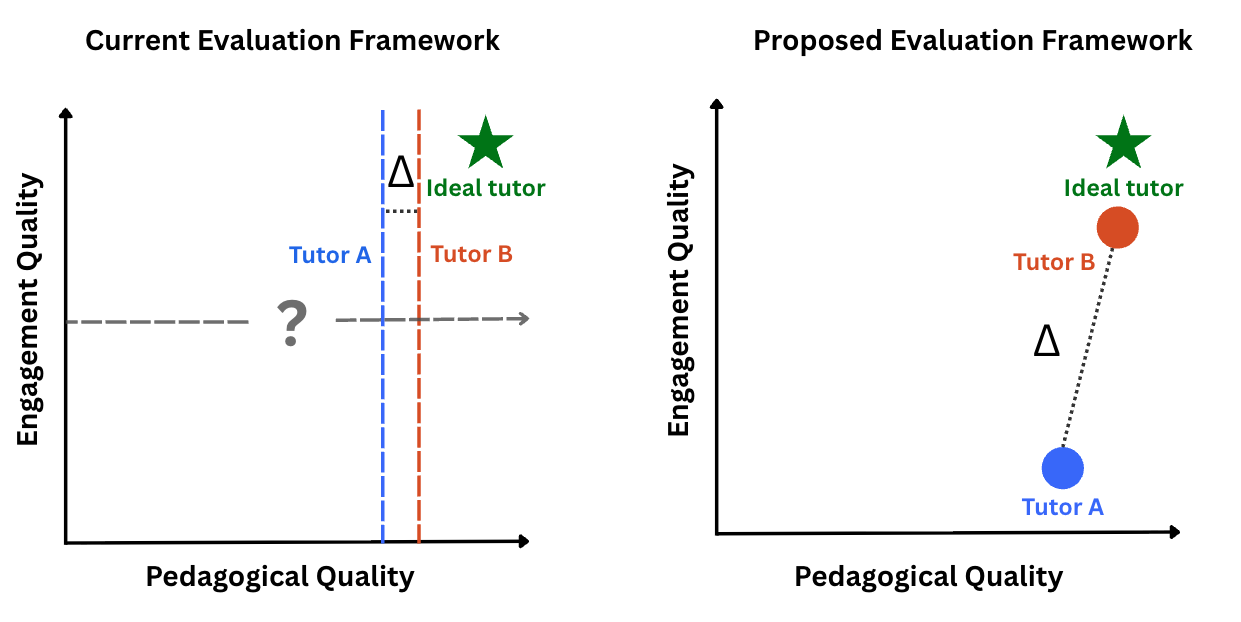}
    \caption{Illustration of the proposed AI tutor evaluation framework. \textbf{Left:} The current evaluation framework positions tutors only along the pedagogical quality axis; as long as pedagogical quality is high, tutors appear comparable in performance. \textbf{Right:} The proposed framework reveals the two-dimensional position of each tutor, capturing both pedagogical quality and student engagement with feedback, providing greater separability between tutors. The star indicates the desired tutor behavior, combining strong pedagogy with effective feedback uptake.}
    \label{fig:contribution}
\end{figure}

As Large Language Models (LLMs)-based tutoring systems---hereafter Artificial Intelligence (AI) tutors---are increasingly deployed to provide students with feedback in classroom settings~\cite{liu2024teaching,liu2025lpitutor,zamfirescu202561a,mittal2025askademia}, it is essential to define how these systems should be evaluated. Existing evaluations primarily assess feedback quality in isolation using predefined pedagogical rubrics~\cite{macina2025mathtutorbench,srinivasa2025tutorbench,maurya-etal-2025-unifying,qi2025knowledge}. While pedagogical quality is a prerequisite for deploying AI tutors, it may not fully capture feedback effectiveness: feedback can be pedagogically sound yet unhelpful if students do not engage with it or struggle to apply it correctly~\cite{sadler1989formative,patchan2016nature,wisniewski2020power}. For example, overly cautious feedback may satisfy pedagogical criteria while failing to support immediate problem-solving, whereas feedback that students readily act upon may violate pedagogical guidelines. These distinctions are not captured by pedagogy-only evaluation, which motivates the need to incorporate students’ engagement with feedback as a complementary evaluation dimension. This need is further amplified by the fact that student–AI tutor interaction fundamentally differs from student–human tutoring~\cite{jiang2025real}: whereas human tutors flexibly adapt pedagogical intent through dialogue and context, AI tutors may interpret pedagogical instructions literally and generate feedback through fixed prompting structures, such that pedagogical intent does not necessarily translate into effective student support~\cite{jurenka2024towards}.

Thus, in this work, we complement pedagogical evaluation with engagement-based metrics that capture whether students act on tutor feedback and whether those actions are applied correctly in the context of introductory programming problem solving. As shown in \autoref{fig:contribution}, by examining both pedagogical quality and students' engagement with feedback, we aim to provide a more comprehensive and actionable framework for evaluating AI tutors deployed in real learning environments. Using data from a large-scale classroom deployment comprising 10,235 code submissions, we show that engagement-based metrics reveal meaningful differences between AI tutors that are not captured by pedagogical quality alone and are more strongly associated with student perception of helpful feedback. We answer the following research questions:

\begin{itemize}
    \item \textbf{RQ1.} How do pedagogical quality and engagement-based behavioral metrics differ in providing separability between AI tutors?
    
    \item \textbf{RQ2.} How do pedagogical quality and engagement-based metrics relate to each other, and to what extent do they capture distinct aspects of tutor effectiveness?
    
    \item \textbf{RQ3.} Which evaluation metrics are more strongly associated with students’ perceived helpfulness of the feedback?
\end{itemize}
\section{Related Work}
\subsection{Evaluation of AI Tutors}
Recent work has introduced evaluation frameworks for AI tutors that assess the quality of tutor feedback along multiple pedagogical dimensions~\cite{gupta2025beyond,tack2022ai,jurenka2024towards}. For example, MathTutorBench~\cite{macina2025mathtutorbench} evaluates open-ended mathematical tutoring across dimensions that include mathematical expertise, support for student understanding, and the quality of instructional scaffolding, using real teacher–student dialogues and learned evaluation models. TutorBench~\cite{srinivasa2025tutorbench} introduces a large-scale dataset of Science, Technology, Engineering, and Mathematics (STEM) tutoring interactions annotated with learning-science–motivated rubrics targeting core tutoring skills such as explanation clarity, adaptation, and misconception diagnosis. MRBench~\cite{maurya-etal-2025-unifying} further operationalizes pedagogical evaluation through expert annotations across eight fine-grained pedagogical dimensions over educational dialogues. TutorGym~\cite{weitekamp2025tutorgym} extends evaluation to interactive settings by embedding AI tutors within simulated tutoring environments. While these evaluation frameworks provide structured static evaluations, they primarily assess tutor responses in isolation, scoring responses based on rubric alignment or pedagogical dimensions and may fail to capture the dynamic and adaptive behavior of generative AI tutors~\cite{macina2023opportunities}. \cite{maurya2025pedagogy} further notes that limitations in reliability, scalability, and unification make it difficult to assess pedagogical effectiveness consistently across educational settings for AI tutors' pedagogy-only assessment. 

\subsection{Behavioral Signals of Feedback Use}

Learning engagement broadly refers to how learners participate in learning activities~\cite{fredricks2004school}. 
Prior work commonly distinguishes between affective, cognitive, and behavioral dimensions of engagement~\cite{fredricks2004school,wong2024student}. 
Affective engagement reflects learners’ emotional responses, cognitive engagement captures the effort and strategies learners apply to a task, and behavioral engagement refers to observable actions during learning activities, such as participation and task completion~\cite{wong2022student,han2019learner}. Although this multidimensional view is well-established, engagement is often treated as a single, loosely defined concept in empirical studies~\cite{henrie2015measuring}. 
Recent work, therefore, argues that engagement should be studied through clearly defined dimensions with explicit operationalization~\cite{wong2024student}. 

In feedback-driven learning settings, behavioral engagement is particularly informative. Feedback is widely understood as an active process in which learners must interpret information and decide how to incorporate it into their work, rather than as passive information delivery~\cite{carless2018development}. Because this process requires learner action, students vary substantially in the extent to which they attend to and use feedback~\cite{jonsson201824}. Consequently, evaluating feedback effectiveness requires examining how students act on feedback in practice, not only whether the feedback itself is well formed~\cite{tay2022students}. 
From this perspective, behavioral engagement reflects how learners translate their intentions into actions during problem solving, which influences performance~\cite{martin2023integrating,miller2015using}.

\section{Methodology}

\subsection{Study Context and Data}
Our study is situated in a large introductory programming course (CS61A~\footnote{https://cs61a.org/}) at the University of California, Berkeley, with an average enrollment of approximately 1,000 students per semester. The course covers core topics including data types, conditionals, loops, functions, recursion, and object-oriented programming. Students complete 10 assignments per semester, each consisting of 3 to 6 programming problems. Students develop their solutions in an online environment with access to an autograder that provides immediate feedback on passed and failed test cases (with no hidden tests). Students may resubmit their solutions without penalty as many times as they would like. 
In addition to autograder feedback, students receive natural-language feedback from an LLM-based AI tutor whenever a submission fails at least one test case. All submission attempts are logged, including the student’s code, the autograder output, and the AI tutor's feedback. 

Our dataset spans two semesters, Fall 2024 and Fall 2025, with 13,569 and 22,363 code submissions, respectively, drawn from the first five assignments focusing on introductory Python concepts. To enable a controlled comparison across tutors, we randomly sample 2 representative problems per assignment that appear in both semesters, yielding 10 problems in total. The sampled data includes 681 students and 3,188 individual code submissions for Fall 2024, and 958 students and 7,047 individual code submissions for Fall 2025, where all submissions for a given student--problem pair are retained as a complete stream rather than sampled independently. On average, students made 3.2 submissions per stream in Fall 2024 and 2.6 in Fall 2025.

\subsection{AI Tutor Variations}

We use two AI tutor configurations, deployed in different semesters of the same course, as case studies to illustrate and validate the proposed evaluation metrics in Section 3.3. Both tutors use the same LLM (GPT-4~\cite{gpt4}) and differ only in their prompting structure.

\begin{itemize}
    \item \textit{\textbf{BaselineTutor}}~\cite{zamfirescu202561a} was deployed in Fall 2024. It generates feedback by conditioning on the problem statement, the student’s current code, the autograder output, and the student’s prior submissions. The system prompt instructs the model to behave as a tutor and produce natural-language feedback to guide the student’s next steps.

    \item \textit{\textbf{MisconceptionTutor}}~\cite{niousha2026misconception} was deployed in Fall 2025. Inspired by the misconception tutoring framework of \cite{ross-andreas-2024-toward}, it extends the baseline tutor with two additional structured steps: (1) Identifying likely student misconceptions from a pre-defined instructor-authored list, and (2) Generating feedback that explicitly targets the identified misconceptions. The model is prompted to output a structured JSON object containing both the detected misconceptions and the corresponding feedback; only the feedback field is shown to students.
\end{itemize}

\subsection{Evaluation Metrics}
\begin{table}[t]
\caption{Pedagogical evaluation dimensions, scoring rubrics, and desired labels, adapted from the unified taxonomy of \cite{maurya-etal-2025-unifying}. For all dimensions, \texttt{DAMR} counts responses whose \textit{Label} matches the \textit{Desired Label} column; for \texttt{tutor\_tone}, we consider both labels 1 (Encouraging) and 2 (Neutral) as desirable.}
\label{tab:ped-dimension}
\centering
\footnotesize
\resizebox{\linewidth}{!}{
\begin{tabular}{
  m{4cm}
  m{5.8cm}
  >{\centering\arraybackslash}m{4.8cm}
  >{\centering\arraybackslash}m{2cm}
}
\hline
\textbf{Dimension} & \textbf{Definition} & \textbf{Labels} & \textbf{Desired Label(s)} \\
\hline
\texttt{mistake\_identification}
& Has the tutor identified or recognized a mistake in the student's response?
& \makecell{1: Yes \\ 2: To some extent \\ 3: No}
& 1 \\ 
\texttt{mistake\_location}
& Does the tutor's response accurately point to a genuine mistake and its location?
& \makecell{1: Yes \\ 2: To some extent \\ 3: No}
& 1 \\
\texttt{revealing\_answer}
& Does the tutor reveal the final answer (whether correct or not)?
& \makecell{1: Yes (correct) \\ 2: Yes (incorrect) \\ 3: No}
& 3 \\
\texttt{providing\_guidance}
& Does the tutor offer correct and relevant guidance, such as an explanation, elaboration, hint, or examples?
& \makecell{1: Yes \\ 2: To some extent \\ 3: No}
& 1 \\
\texttt{actionability}
& Is it clear from the tutor's feedback what the student should do next?
& \makecell{1: Yes \\ 2: To some extent \\ 3: No}
& 1 \\
\texttt{coherence}
& Is the tutor's response logically consistent with the student's previous responses?
& \makecell{1: Yes \\ 2: To some extent \\ 3: No}
& 1 \\
\texttt{tutor\_tone}
& Is the tutor's response encouraging, neutral, or offensive?
& \makecell{1: Encouraging \\ 2: Neutral \\ 3: Offensive}
& 1, 2 \\
\texttt{humanness}
& Does the tutor's response sound natural rather than robotic or artificial?
& \makecell{1: Yes \\ 2: To some extent \\ 3: No}
& 1 \\
\hline
\end{tabular}
}
\end{table}
\label{sec:method-eval}
\subsubsection{Pedagogy-based Evaluation}
To evaluate the AI tutors from a pedagogical perspective, we adopt an evaluation framework by Maurya et al. \cite{maurya-etal-2025-unifying} that assesses the \emph{pedagogical quality} of AI tutor feedback. The taxonomy operationalizes pedagogical quality along eight learning-science–motivated dimensions. \autoref{tab:ped-dimension} summarizes these dimensions and their desired criteria. 

Following prior work~\cite{maurya-etal-2025-unifying}, we summarize pedagogical quality using the proposed metric,
\textbf{Desired Annotation Match Rate (\texttt{DAMR})}. For a given AI tutor and
dimension, \texttt{DAMR} is defined as:
\[
\texttt{DAMR} = \frac{1}{N} \sum_{i=1}^{N} \mathbb{I}[\ell_i = \ell^{*}]
\]
where \(N\) is the number of evaluated feedback messages, \(\ell_i\) is the annotation
label for the \(i\)-th feedback on that dimension, \(\ell^{*}\) is the desired label,
and \(\mathbb{I}[\cdot]\) is an indicator function equal to 1 when the assigned label matches the desired label and 0 otherwise. We use GPT-4.1~\cite{gpt41} (version: 2025-04-14, temperature=0.0) to assign labels to each feedback message according to the pedagogical taxonomy and scoring rubrics in \autoref{tab:ped-dimension} and the prompt used in their work. 
On 20 sampled interactions annotated by two humans, average Cohen’s $\kappa$ across dimensions indicates substantial human--human agreement ($\kappa=0.76$) and moderate LLM--human agreement ($\kappa=0.65$, $0.44$).

\subsubsection{Engagement-based Evaluation.}
We quantify the \textit{engagement quality} of tutor feedback by examining \textbf{whether} and \textbf{how} students used the AI tutor's feedback, conditioned on their subsequent code revisions.

We evaluate AI tutors along two dimensions of engagement quality: \textbf{relevance} (whether feedback is used) and \textbf{success} (whether used feedback is applied correctly).


For a given student--problem pair $(s_i, p_u)$, we denote two consecutive student submissions at times $t$ and $t+1$ as $c_{t, s_i, p_u}$ and $c_{t+1, s_i, p_u}$, respectively. Each tutor feedback message $f_{t, s_i, p_u}$ consists of $M_{t, s_i, p_u}$ sentences $\{\ell_{t,1}, \ldots, \ell_{t,M_{t, s_i, p_u}}\}$.
For each sentence \(\ell_{t,j}\), an LLM judge (GPT-4.1, version: 2025-04-14, temperature=0.0) assigns attribution labels
\[
\text{\texttt{rel}}(c_{t, s_i, p_u}, \ell_{t,j}, c_{t+1, s_i, p_u}) \in \{0,1\}
\]
\[
\text{\texttt{succ}}(c_{t, s_i, p_u}, \ell_{t,j}, c_{t+1, s_i, p_u}) \in \{0,1\}
\]
indicating whether the sentence influenced the student’s code edit (relevance or \texttt{rel}) and, when relevant, whether the suggested change was applied correctly (success or \texttt{succ}). Success is evaluated only when \(\texttt{rel}=1\). The judge also produces a rationale grounded in the student’s code edits. On 20 randomly sampled interactions annotated by two humans, LLM--human agreement yields $\kappa=0.67$--$0.76$ for \texttt{rel} and $0.80$--$1.00$ for \texttt{succ}, with human--human agreement of $\kappa=0.89$ (\texttt{rel}) and $\kappa=0.80$ (\texttt{succ}), indicating substantial agreement.


For each feedback message, we define two engagement quality scores:
\[
\text{\texttt{\texttt{RelScore}}}_{f_{ t, s_i, p_u}} = \frac{1}{M_{t, s_i, p_u}} \sum_{j=1}^{M_{t, s_i, p_u}} \text{\texttt{rel}}(c_{t, s_i, p_u}, \ell_{t,j}, c_{t+1, s_i, p_u})
\]
\[
\text{\texttt{SuccScore}}_{f_{ t, s_i, p_u}} =
\frac{ \sum_{j=1}^{M_{t, s_i, p_u}} \text{\texttt{succ}}(c_{t, s_i, p_u}, \ell_{t,j}, c_{t+1, s_i, p_u}) }
     { \sum_{j=1}^{M_{t, s_i, p_u}} \text{\texttt{rel}}(c_{t, s_i, p_u}, \ell_{t,j}, c_{t+1, s_i, p_u}) }
\]
where $\text{\texttt{RelScore}}_{f_{t, s_i, p_u}}$ measures the fraction of feedback sentences that student $s_i$ engages with based on feedback $f_{t, s_i, p_u}$, and $\text{\texttt{SuccScore}}_{f_{ t, s_i, p_u}}$ measures the fraction of engaged feedback that is applied correctly.


\subsection{Predicting Student Perception on Helpfulness}


To ground our evaluation in the student perspective, we collect students' self-reported perceptions of feedback helpfulness. We do not claim that perceived helpfulness reflects learning outcomes, but rather use it as a validation signal to assess whether pedagogical and engagement-based metrics align with how students experience feedback in practice. After each failed submission, students were optionally asked to rate the AI tutor’s response on a 5-point Likert scale with the following options: \textbf{5} = helpful, all fixed; \textbf{4} = helpful, not all fixed; \textbf{3} = not helpful but made sense; \textbf{2} = not helpful and did not make sense; and \textbf{1} = misleading or wrong feedback. Student ratings were available for approximately 38\% of sampled submissions. Specifically, among the sampled submissions, we received 954 feedback ratings (from 274 students) under \textit{BaselineTutor} (Fall 2024), and 2,915 ratings (from 639 students) under \textit{MisconceptionTutor} (Fall 2025).


We map students’ Likert ratings to a binary outcome, where ratings of 4--5 are coded as 1 and ratings of 1--3 are coded as 0. For each feedback message, we represent pedagogical quality using binary indicators
$P_1, P_2, \ldots, P_8$, one for each pedagogical dimension, where
$P_j = 1$ if the $j$-th pedagogical dimension meets its desired criterion
and $P_j = 0$ otherwise. We also compute engagement quality scores $\texttt{RelScore}$ and $\texttt{SuccScore}$.

We fit the following three binary logistic regression models to disentangle the effects of pedagogical and engagement-based metrics on students’ perceived helpfulness of the feedback message:
\begin{enumerate}
    \item \textbf{Pedagogy-only model}, with covariates
    \[
    \mathbf{x} = \bigl[P_1, P_2, \ldots, P_8,\ \mathbb{I}[\textit{BaselineTutor}]\bigr]
    \]
    where $P_j$ is a binary indicator for whether the $j$-th pedagogical dimension meets its desired criterion and $\mathbb{I}[\textit{BaselineTutor}]$ is a binary semester control equal to 1 for \textit{BaselineTutor} (Fall~2024) and 0 for \textit{MisconceptionTutor} (Fall~2025).

    \item \textbf{Engagement-only model}, with covariates
    \[
    \mathbf{x} = \bigl[\texttt{RelScore},\ \texttt{SuccScore},\ \mathbb{I}[\textit{BaselineTutor}]\bigr].
    \]

    \item \textbf{Combined model}, with covariates
    \[
    \mathbf{x} = \bigl[\texttt{RelScore},\ \texttt{SuccScore},\ P_1, P_2, \ldots, P_8,\ \mathbb{I}[\textit{BaselineTutor}]\bigr].
    \]
\end{enumerate}

Each model estimates the probability of student-perceived helpfulness as
\[
\Pr(y = 1) = \sigma(\mathbf{x}^\top \boldsymbol{\beta}),
\]
where $\mathbf{x}$ denotes the covariates corresponding to each model specification.
Coefficients are interpreted as log-odds of helpfulness, holding all other variables constant.


\section{Results}

\begin{table}[t]
\centering
\caption{Pedagogical quality comparison between \textit{BaselineTutor} (Fall~2024) and \textit{MisconceptionTutor} (Fall~2025) using \texttt{DAMR}. Bolded values indicate the higher \texttt{DAMR} (or ties) between AI tutors for each dimension. Differences are evaluated using Fisher’s exact test with Holm-corrected
$p$-values ($^{*}p<0.05$), with Cohen’s $h$ reported as the effect size.}

\label{tab:damr}
\small
\begin{tabular}{lcccc}
\toprule
\textbf{Dimension} &
\textbf{BaselineTutor} &
\textbf{MisconceptionTutor} &
\textbf{Cohen's $h$} &
\textbf{$p$} \\
\midrule
\texttt{mistake\_identification} & 94.20 & \textbf{98.20} & 0.217 & $<.001^{*}$ \\
\texttt{mistake\_location}      & 89.96 & \textbf{95.81} & 0.233 & $<.001^{*}$ \\
\texttt{revealing\_answer}       & 93.95 & \textbf{99.65} & 0.378 & $<.001^{*}$ \\
\texttt{providing\_guidance}     & 88.96 & \textbf{89.85} & 0.029 & 0.340 \\
\texttt{actionability}         & \textbf{89.15} & 87.99 & $-0.036$ & 0.278 \\
\texttt{coherence}              & 95.61 & \textbf{98.69} & 0.193 & $<.001^{*}$ \\
\texttt{tutor\_tone}             & \textbf{100.00} & \textbf{100.00} & 0.000 & 1.000 \\
\texttt{humanness}              & 99.56 & \textbf{99.94} & 0.085 & $<.001^{*}$ \\
\bottomrule
\end{tabular}

\vspace{0.5em}
\end{table}

\begin{figure}[t]
    \centering
    \includegraphics[width=0.95\linewidth]{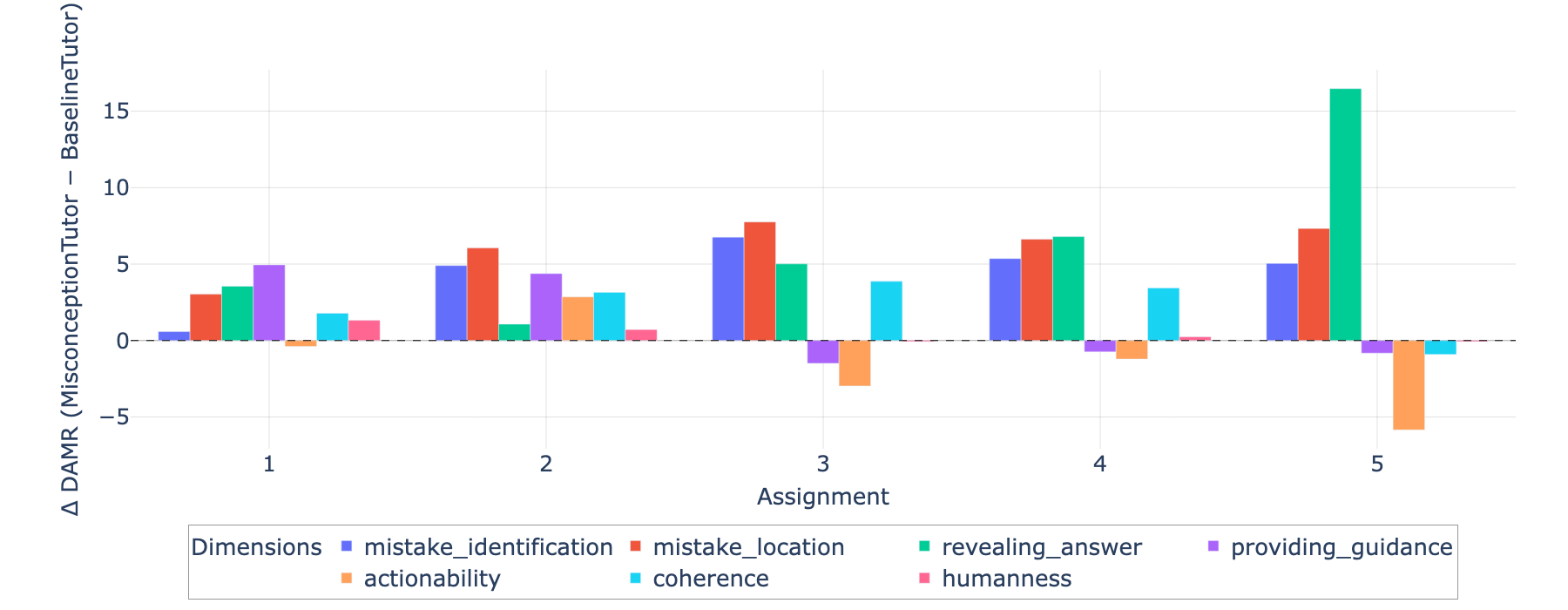}
    \caption{Per-assignment differences in \texttt{DAMR} between \textit{MisconceptionTutor} and \textit{BaselineTutor}. Positive values indicate higher \texttt{DAMR} for \textit{MisconceptionTutor}. \texttt{tutor\_tone} is omitted, as it shows no difference between tutors, with 100\% \texttt{DAMR} for both.}
    \label{fig:damr_per_hw}
\end{figure}

\begin{table}[t]
\centering
\small
\caption{Per-assignment feedback engagement quality comparison between \textit{BaselineTutor} (Fall~2024) and \textit{MisconceptionTutor} (Fall~2025). Reported values are the mean $\pm$ standard deviation (\%) of feedback relevance score (\texttt{RelScore}) and success score (\texttt{SuccScore}). Bolded values indicate the higher mean between tutors for each assignment. Statistical significance is assessed using Mann--Whitney U tests ($^{*}p<0.05$), with Holm-corrected $p$-values across assignments within each engagement metric.}
\label{tab:sentence_engagement_per_assignment}
\begin{tabular}{ccccc}
\toprule
\textbf{Assignment} &
\textbf{Metric} &
\textbf{BaselineTutor} &
\textbf{MisconceptionTutor} &
\textbf{$p$} \\
\midrule
1 & \texttt{RelScore} & $67.3 \pm 40.1$ & $\mathbf{82.5 \pm 32.2}$ & $<.001^{*}$ \\
2 & \texttt{RelScore} & $71.7 \pm 31.9$ & $\mathbf{80.9 \pm 28.1}$ & $<.001^{*}$ \\
3 & \texttt{RelScore} & $66.7 \pm 34.2$ & $\mathbf{82.4 \pm 26.5}$ & $<.001^{*}$ \\
4 & \texttt{RelScore} & $73.8 \pm 31.0$ & $\mathbf{86.3 \pm 23.8}$ & $<.001^{*}$ \\
5 & \texttt{RelScore} & $64.0 \pm 31.7$ & $\mathbf{84.9 \pm 27.0}$ & $<.001^{*}$ \\
\midrule
1 & \texttt{SuccScore} & $26.3 \pm 41.2$ & $\mathbf{56.0 \pm 47.9}$ & $<.001^{*}$ \\
2 & \texttt{SuccScore} & $32.6 \pm 41.6$ & $\mathbf{51.6 \pm 45.9}$ & $<.001^{*}$ \\
3 & \texttt{SuccScore} & $49.0 \pm 44.4$ & $\mathbf{51.7 \pm 45.8}$ & 0.469 \\
4 & \texttt{SuccScore} & $\mathbf{60.2 \pm 44.3}$ & $55.6 \pm 45.3$ & $0.035^{*}$ \\
5 & \texttt{SuccScore} & $47.7 \pm 43.1$ & $\mathbf{52.7 \pm 46.5}$ & 0.459 \\
\bottomrule
\end{tabular}
\end{table}

\begin{figure}[t]
    \centering
    \includegraphics[width=0.88\linewidth]{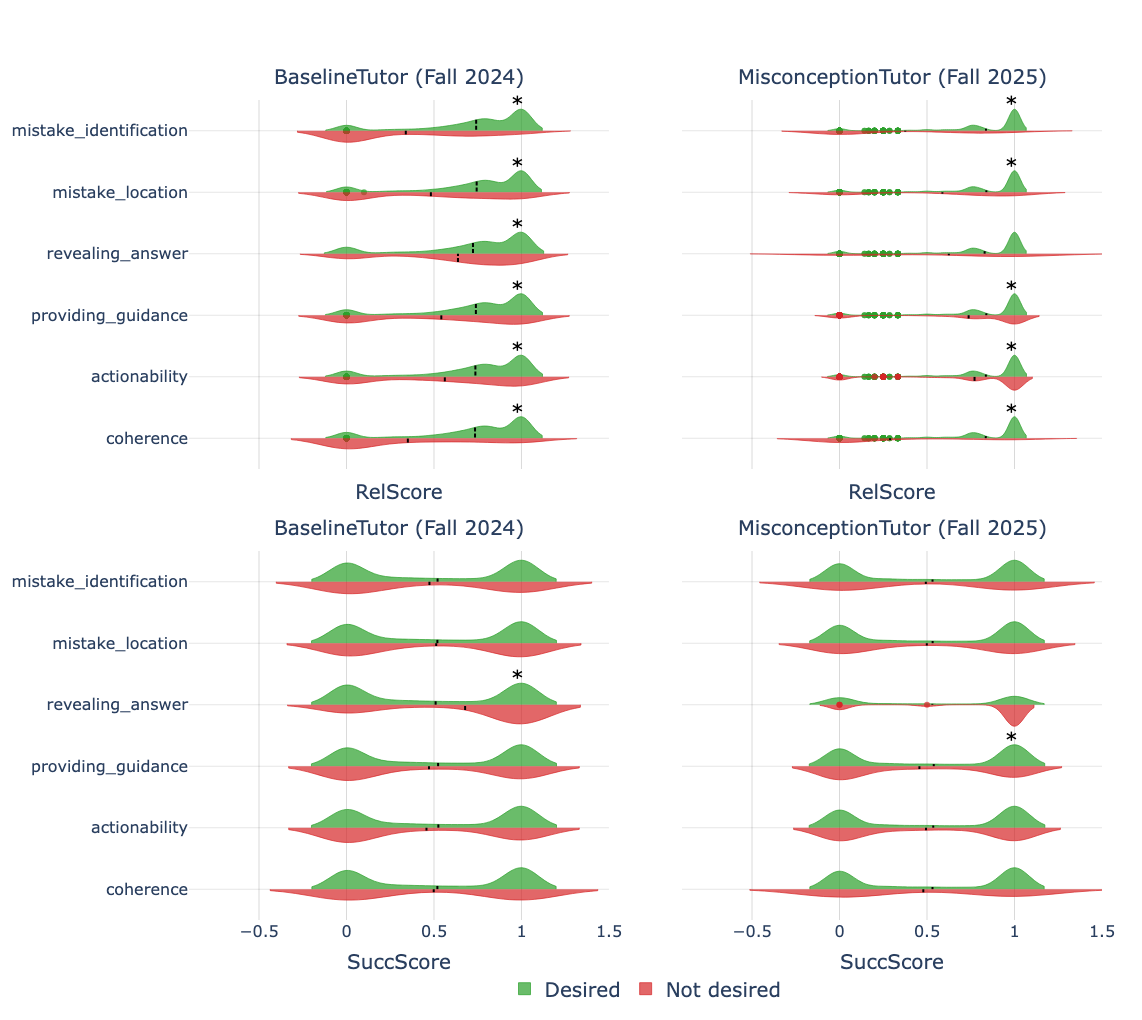}
    \caption{Distribution of \texttt{RelScore} and \texttt{SuccScore} for desired (green) versus undesired (red) feedback messages, shown separately for \textit{BaselineTutor} (Fall 2024) and \textit{MisconceptionTutor} (Fall 2025). Violin widths are scaled within each pedagogical dimension and within each subplot to enable direct comparison between desired and undesired feedback without confounding differences in sample size across dimensions. Black lines indicate means; stars mark statistically significant differences between desired and undesired feedback after Holm–Bonferroni–corrected two-sided Mann–Whitney U tests ($p<0.05$). The \texttt{tutor\_tone} and \texttt{humanness} dimensions are excluded due to insufficient sample sizes in the undesired feedback category ($n < 15$).}
    \label{fig:ped_vs_engage}
    \vspace{-1em}
\end{figure}

\subsection{RQ1: Pedagogical Quality of the AI Tutors}

\autoref{tab:damr} summarizes the pedagogical quality of  \textit{BaselineTutor} (Fall~2024) and \textit{MisconceptionTutor} (Fall~2025) using \texttt{DAMR}. \textit{MisconceptionTutor} achieves higher scores on nearly all dimensions and assignments. The majority of differences are statistically significant. Notably, \textit{MisconceptionTutor} adopts a more conservative feedback strategy and reduces answer revealing compared to \textit{BaselineTutor} (\texttt{DAMR}: 99.65 vs.\ 93.95; $p<.001$). However, this improvement coincides with a decrease in actionability (87.99 vs.\ 89.15), suggesting that more high-level, misconception-focused feedback trades off immediate actionability. Across all dimensions, effect sizes are generally small to moderate, indicating that the \textit{MisconceptionTutor} improves pedagogical quality consistently yet incrementally rather than through large shifts. Moreover, \autoref{fig:damr_per_hw} demonstrates the difference between the AI tutors by plotting per-assignment differences in \texttt{DAMR} between \textit{MisconceptionTutor} and \textit{BaselineTutor} across pedagogical dimensions. Positive values indicate higher \texttt{DAMR} for \textit{MisconceptionTutor}. Across all assignments, most dimensions favor \textit{MisconceptionTutor}, with larger gaps between the two AI tutors on later, more conceptually difficult assignments. In particular, gaps in \texttt{mistake\_identification}, \texttt{mistake\_location}, and reduced \texttt{revealing\_answer} widen over time. Conversely, \texttt{actionability} decreases on later assignments for \textit{MisconceptionTutor}, indicating more conservative feedback.

\subsection{RQ1: Engagement Quality of the AI Tutors}
\autoref{tab:sentence_engagement_per_assignment} reports engagement quality scores, measured by relevance (\texttt{RelScore}) and success (\texttt{SuccScore}), aggregated per assignment. Across all assignments, \textit{MisconceptionTutor} achieves substantially higher \texttt{RelScore}, improving relevance by 9–21 percentage points on every assignment. This indicates that a consistently larger fraction of its feedback influences students’ subsequent code edits. These differences are statistically significant for every assignment. Improvements in \texttt{SuccScore} are more mixed. \textit{MisconceptionTutor} substantially improves success on earlier assignments, but differences diminish or reverse on later assignments, with no significant improvement on Assignments~3 and~5. This suggests that while misconception-focused feedback is more likely to engage students, it does not always translate into immediate correct application on more challenging assignments. Comparing the engagement quality of the two AI tutors introduces a complementary evaluation dimension that is not captured by pedagogical quality alone. 

\subsection{RQ2: Relationship between Pedagogy and Engagement}
Next, we examine how the pedagogical quality, measured by \texttt{DAMR}, translates into student engagement quality with AI tutor feedback. Specifically, we evaluate whether feedback that satisfies pedagogical criteria is more likely to be engaged with by students (high \texttt{RelScore}) and more likely to be applied correctly (high \texttt{SuccScore}). \autoref{fig:ped_vs_engage} shows the distribution of \texttt{RelScore} and \texttt{SuccScore} for feedback that satisfies the desired criteria versus feedback that does not, stratified by pedagogical dimension and AI tutor. Across both AI tutors, pedagogically desired feedback consistently exhibits a higher mean \texttt{RelScore} across all dimensions, with most differences being statistically significant. The full distributions are shifted toward higher \texttt{RelScore} values, indicating that pedagogical dimensions are behaviorally meaningful for student engagement. In contrast, \texttt{SuccScore} distributions for desired and undesired feedback largely overlap for most dimensions, with only modest differences in central tendency, which suggests that pedagogical quality primarily influences which feedback students choose to act on rather than whether those actions lead to correct edits by students. A notable exception is \texttt{revealing\_answer}, which is the only dimension in which the pattern of pedagogically desired feedback achieving higher scores is reversed across both semesters. For \textit{BaselineTutor}, feedback marked as undesired in \texttt{revealing\_answer} achieves a higher mean \texttt{SuccScore} (67.8\%) than desired feedback (50.9\%), and a similar pattern appears in \textit{MisconceptionTutor} (79.4\% vs.\ 53.0\%), with a strong concentration of high \texttt{SuccScore} for undesired feedback. This observation suggests that these successes are driven by students copying the revealed answer rather than understanding and solving the problem themselves. These results underscore the need to interpret engagement metrics alongside pedagogical quality, as high \texttt{SuccScore} alone does not necessarily reflect effective learning, given that success can also be achieved through answer revelation.

\subsection{RQ3: Predictors of Student Satisfaction}
\autoref{tab:binary_class} reports three binary logistic regression models predicting whether students rated AI tutor feedback as highly helpful: a pedagogy-only model, an engagement-only model, and a combined model. Positive coefficients indicate an increase in the log-odds of the outcome, indicating that the feature is associated with higher perceived helpfulness. Across models, engagement-based metrics are the most robust predictors of perceived student helpfulness on a feedback message. In the engagement-only model, both \texttt{RelScore} ($\beta=0.420$, $p<0.001$) and \texttt{SuccScore} ($\beta=0.187$, $p<0.001$) are positively and significantly associated with higher perceived helpfulness. These effects remain stable in the combined model, with nearly identical coefficient magnitudes, indicating that engagement explains variance in students' perceived helpfulness of the feedback message. In contrast, most pedagogical dimensions show weak or inconsistent associations with perceived helpfulness. In the combined model, \texttt{mistake\_identification} shows a significantly negative association ($\beta=-0.524$, $p=0.038$), whereas \texttt{providing\_guidance} shows a significantly positive association ($\beta=0.349$, $p=0.019$). This result suggests that students find feedback helpful when it guides them on the next step, and less helpful when it only identifies their mistake. Finally, the negative coefficient for \textit{BaselineTutor} across all models indicates consistently lower perceived helpfulness relative to \textit{MisconceptionTutor}. Furthermore, low pseudo-R$^2$ values are expected for subjective ratings influenced by unobserved factors (e.g., prior knowledge, task difficulty); thus, we focus on relative explanatory power rather than absolute prediction.

\begin{table}[t]
\centering
\caption{Binary logistic regression results predicting high perceived helpfulness of AI tutor feedback. Columns correspond to three model specifications: a pedagogy-only model using binary indicators for pedagogical quality, an engagement-only model using \texttt{RelScore} and \texttt{SuccScore}, and a combined model including all attributes. Entries report log-odds coefficients with standard errors in parentheses; significance is indicated by $^{*}p<0.05$, $^{**}p<0.01$, and $^{***}p<0.001$. All models include a binary semester control (\textit{BaselineTutor}, Fall~2024 vs.\ Fall~2025). Pseudo $R^2$ reflects improvement over a null model; we focus on relative coefficient magnitudes and robustness across model specifications. The \texttt{tutor\_tone} dimension is omitted from the models because all feedback messages were labeled as pedagogically desired, resulting in no variance.}

\label{tab:binary_class}
\small
\begin{tabular}{llll}
\toprule
 & \multicolumn{3}{c}{\textbf{Model}} \\
\cmidrule(lr){2-4}
 & Pedagogy-only & Engagement-only & Combined \\
\midrule

\texttt{mistake\_identification} 
& $-$0.155 (0.211) 
& -- 
& $-$0.524$^{*}$ (0.253) \\

\texttt{mistake\_location} 
& 0.135 (0.145) 
& -- 
& 0.260 (0.161) \\

\texttt{revealing\_answer} 
& $-$0.203 (0.152) 
& -- 
& $-$0.206 (0.167) \\

\texttt{providing\_guidance} 
& 0.444$^{**}$ (0.139) 
& -- 
& 0.349$^{*}$ (0.148) \\

\texttt{actionability} 
& $-$0.179 (0.125) 
& -- 
& $-$0.097 (0.133) \\

\texttt{coherence} 
& $-$0.187 (0.219) 
& -- 
& $-$0.111 (0.288) \\

\texttt{humanness} 
& 1.093 (0.760) 
& -- 
& 1.261 (1.072) \\

\texttt{RelScore} 
& -- 
& 0.420$^{***}$ (0.127) 
& 0.434$^{***}$ (0.128) \\

\texttt{SuccScore} 
& -- 
& 0.187$^{***}$ (0.049) 
& 0.176$^{***}$ (0.049) \\

\midrule
\textit{BaselineTutor}
& $-$0.297$^{***}$ (0.049) 
& $-$0.226$^{***}$ (0.051) 
& $-$0.237$^{***}$ (0.052) \\

Pseudo $R^2$ 
& 0.0048 & 0.0047 & 0.0064 \\
\bottomrule
\end{tabular}

\end{table}

\section{Discussion}





  
\textbf{Pedagogical quality is necessary but not sufficient.} Pedagogy-based metrics function as a vital prerequisite for AI tutor deployment: they establish a bar for instructional quality and appropriateness, allowing diagnosis of pedagogically undesired feedback. However, once AI tutors reach a high pedagogical standard, rubric-based metrics provide less clear differentiation between systems, suggesting that pedagogy-based evaluation alone may be insufficient, especially at higher performance levels.


\textbf{Engagement provides an additional and behaviorally grounded dimension of separability.} Engagement-based metrics distinguish AI tutors in a different way than pedagogical metrics. In our results, \textit{MisconceptionTutor} consistently achieves higher feedback relevance across all assignments, indicating that a larger fraction of its feedback influences students’ subsequent code edits, a pattern not reflected in pedagogical scores alone. Moreover, engagement-based measures are more strongly associated with students' perception of feedback helpfulness than pedagogy-only metrics.

\textbf{Effective AI tutors require balanced evaluation.} Optimizing solely for pedagogical quality risks producing feedback that is technically sound but insufficiently engaging, while optimizing for engagement alone can encourage superficial success. Pedagogical quality has its greatest impact when students actively engage with the feedback. Our results illustrate this balance: \textit{MisconceptionTutor} generates substantially more engaging feedback across all assignments and improves the correctness of students' applications on earlier assignments, despite being more pedagogically conservative. This trade-off highlights the need for evaluation frameworks that jointly consider pedagogical quality and student engagement.

\section{Limitations and Future Work}

Our engagement-based metrics measure immediate feedback uptake but do not assess longer-term learning outcomes; future work should relate these metrics to learning gains measured through post-tests or transfer tasks. Next, because our comparison spans two semesters, population differences may confound observed effects; future work will validate the framework within a single semester using randomized A/B deployments to isolate tutor effects. While our framework is instantiated in a programming context, it can generalize to other settings as long as a directional success criterion can be defined --- for instance, in dialog-based tutoring, \texttt{SuccScore} could capture whether a student's next response moves toward the desired understanding. Additionally, although LLM-human agreement for pedagogical dimensions is moderate, this is consistent with prior work highlighting the inherent subjectivity of pedagogical quality assessment~\cite{maurya2025pedagogy}, and LLM-based annotation remains a practical necessity for scaling evaluation. Finally, \texttt{RelScore} and \texttt{SuccScore} measure alignment between feedback and code edits rather than direct feedback reading, making it difficult to distinguish revisions not driven by the feedback; edge cases such as large-scale rewrites or partially adopted suggestions may further complicate attribution, warranting finer-grained scoring in future work.
\section{Conclusion}
We show that pedagogical quality is effective for evaluating whether AI tutors meet instructional standards, but insufficient for capturing how students respond to feedback in practice, an important dimension of feedback effectiveness. We introduce an engagement-based evaluation framework grounded in students’ revisions to assess whether and how feedback is used. Across two classroom deployments, these behavioral signals reveal differences between tutors not captured by pedagogy-only metrics and are more strongly associated with students’ perceived helpfulness of feedback. Together, our results highlight the need to evaluate AI tutors along both pedagogical and engagement dimensions.

\bibliography{main}

@inproceedings{qi2025knowledge,
  title={A knowledge-component-based methodology for evaluating ai assistants},
  author={Qi, Laryn and Zamfirescu-Pereira, JD and Kim, Taehan and Hartmann, Bj{\"o}rn and DeNero, John and Norouzi, Narges},
  booktitle={Proceedings of the ACM Global on Computing Education Conference 2025 Vol 1},
  year={2025}
}

@inproceedings{mittal2025askademia,
  title={Askademia: A real-time {AI} system for automatic responses to student questions},
  author={Mittal, Meenakshi and Tyagi, Gaurav and Bailey, Azalea and Ranade, Gireeja and Norouzi, Narges},
  booktitle={International Conference on Artificial Intelligence in Education},
  year={2025},
  organization={Springer}
}

@inproceedings{ross-andreas-2024-toward,
    title = "Toward In-Context Teaching: Adapting Examples to Students' Misconceptions",
    author = "Ross, Alexis  and
      Andreas, Jacob",
    booktitle = "Proceedings of the 62nd Annual Meeting of the Association for Computational Linguistics (Volume 1: Long Papers)",
    year = "2024",
}

@inproceedings{macina2025mathtutorbench,
  title={Mathtutorbench: A benchmark for measuring open-ended pedagogical capabilities of {LLM} tutors},
  author={Macina, Jakub and Daheim, Nico and Hakimi, Ido and Kapur, Manu and Gurevych, Iryna and Sachan, Mrinmaya},
  booktitle={Proceedings of the 2025 Conference on Empirical Methods in Natural Language Processing},
  year={2025}
}

@article{srinivasa2025tutorbench,
  title={TutorBench: A Benchmark To Assess Tutoring Capabilities Of Large Language Models},
  author={Srinivasa, Rakshith S and Che, Zora and Zhang, Chen Bo Calvin and Mares, Diego and Hernandez, Ernesto and Park, Jayeon and Lee, Dean and Mangialardi, Guillermo and Ng, Charmaine and Cardona, Ed-Yeremai Hernandez and others},
  journal={arXiv preprint arXiv:2510.02663},
  year={2025}
}

@inproceedings{weitekamp2025tutorgym,
  title={TutorGym: A Testbed for Evaluating {AI} Agents as Tutors and Students},
  author={Weitekamp, Daniel and N. Siddiqui, Momin and J. MacLellan, Christopher},
  booktitle={International Conference on Artificial Intelligence in Education},
  year={2025},
  organization={Springer}
}

@inproceedings{maurya-etal-2025-unifying,
    title = "Unifying {AI} Tutor Evaluation: An Evaluation Taxonomy for Pedagogical Ability Assessment of {LLM}-Powered {AI} Tutors",
    author = "Maurya, Kaushal Kumar  and
      Srivatsa, Kv Aditya  and
      Petukhova, Kseniia  and
      Kochmar, Ekaterina",
    editor = "Chiruzzo, Luis  and
      Ritter, Alan  and
      Wang, Lu",
    booktitle = "Proceedings of the 2025 Conference of the Nations of the Americas Chapter of the Association for Computational Linguistics: Human Language Technologies (Volume 1: Long Papers)",
    month = apr,
    year = "2025",
    address = "Albuquerque, New Mexico",
    publisher = "Association for Computational Linguistics",
    ISBN = "979-8-89176-189-6"
}

@misc{gpt41,
    title={Introducing {GPT-4.1} in the {API}},
    author={OpenAI},
    year={2025},
}

@misc{gpt4,
    title={{GPT-4}},
    author={OpenAI},
    year={2023},
}

@inproceedings{gupta2025beyond,
  title={Beyond final answers: Evaluating large language models for math tutoring},
  author={Gupta, Adit and Reddig, Jennifer and Calo, Tommaso and Weitekamp, Daniel and MacLellan, Christopher J},
  booktitle={International Conference on Artificial Intelligence in Education},
  year={2025},
  organization={Springer}
}

@article{jurenka2024towards,
  title={Towards responsible development of generative {AI} for education: An evaluation-driven approach},
  author={Jurenka, Irina and Kunesch, Markus and McKee, Kevin R and Gillick, Daniel and Zhu, Shaojian and Wiltberger, Sara and Phal, Shubham Milind and Hermann, Katherine and Kasenberg, Daniel and Bhoopchand, Avishkar and others},
  journal={arXiv preprint arXiv:2407.12687},
  year={2024}
}

@inproceedings{maurya2025pedagogy,
  title={Pedagogy-Driven Evaluation of Generative {AI-Powered} Intelligent Tutoring Systems},
  author={Maurya, Kaushal Kumar and Kochmar, Ekaterina},
  booktitle={International Conference on Artificial Intelligence in Education},
  year={2025},
  organization={Springer}
}

@inproceedings{tack2022ai,
 address = {Durham, United Kingdom},
 author = {AnaÃ¯s Tack and Chris Piech},
 booktitle = {Proceedings of the 15th International Conference on Educational Data Mining},
 isbn = {978-1-7336736-3-1},
 month = {July},
 publisher = {International Educational Data Mining Society},
 title = {The {AI} Teacher Test: Measuring the Pedagogical Ability of Blender and {GPT-3} in Educational Dialogues},
 year = {2022}
}

@inproceedings{zamfirescu202561a,
  title={{61A} Bot Report: {AI} Assistants in {CS1} Save Students Homework Time and Reduce Demands on Staff. ({Now} What?)},
  author={Zamfirescu-Pereira, JD and Qi, Laryn and Hartmann, Bj{\"o}rn and DeNero, John and Norouzi, Narges},
  booktitle={Proceedings of the 56th ACM Technical Symposium on Computer Science Education V. 1},
  year={2025}
}

@inproceedings{liu2024teaching,
  title={Teaching {CS50 with AI}: leveraging generative artificial intelligence in computer science education},
  author={Liu, Rongxin and Zenke, Carter and Liu, Charlie and Holmes, Andrew and Thornton, Patrick and Malan, David J},
  booktitle={Proceedings of the 55th ACM technical symposium on computer science education V. 1},
  year={2024}
}

@article{liu2025lpitutor,
  title={{LPITutor: an LLM} based personalized intelligent tutoring system using {RAG} and prompt engineering},
  author={Liu, Zhensheng and Agrawal, Prateek and Singhal, Saurabh and Madaan, Vishu and Kumar, Mohit and Verma, Pawan Kumar},
  journal={PeerJ Computer Science},
  volume={11},
  year={2025},
  publisher={PeerJ Inc.}
}

@article{patchan2016nature,
  title={The nature of feedback: How peer feedback features affect students’ implementation rate and quality of revisions.},
  author={Patchan, Melissa M and Schunn, Christian D and Correnti, Richard J},
  journal={Journal of Educational Psychology},
  volume={108},
  number={8},
  pages={1098},
  year={2016},
  publisher={American Psychological Association}
}

@inproceedings{macina2023opportunities,
    title = "Opportunities and Challenges in Neural Dialog Tutoring",
    author = "Macina, Jakub  and
      Daheim, Nico  and
      Wang, Lingzhi  and
      Sinha, Tanmay  and
      Kapur, Manu  and
      Gurevych, Iryna  and
      Sachan, Mrinmaya",
    editor = "Vlachos, Andreas  and
      Augenstein, Isabelle",
    booktitle = "Proceedings of the 17th Conference of the European Chapter of the Association for Computational Linguistics",
    month = may,
    year = "2023",
    address = "Dubrovnik, Croatia",
    publisher = "Association for Computational Linguistics",
}

@article{sadler1989formative,
  title={Formative assessment and the design of instructional systems},
  author={Sadler, D Royce},
  journal={Instructional science},
  volume={18},
  number={2},
  year={1989},
  publisher={Springer}
}

@article{wisniewski2020power,
  title={The power of feedback revisited: A meta-analysis of educational feedback research},
  author={Wisniewski, Benedikt and Zierer, Klaus and Hattie, John},
  journal={Frontiers in psychology},
  volume={10},
  year={2020},
  publisher={Frontiers}
}

@inproceedings{jiang2025real,
  title={How Real Is {AI} Tutoring? {Comparing} Simulated and Human Dialogues in One-on-One Instruction},
  author={Li, Ruijia and Jiang, Yuan-Hao and Wang, Jiatong and Jiang, Bo},
  booktitle={International Conference on Computers in Education},
  year={2025}
}

@incollection{jonsson201824, place={Cambridge}, series={Cambridge Handbooks in Psychology}, title={Facilitating Students’ Active Engagement with Feedback}, booktitle={The Cambridge Handbook of Instructional Feedback}, publisher={Cambridge University Press}, author={Jonsson, Anders and Panadero, Ernesto}, editor={Lipnevich, Anastasiya A. and Smith, Jeffrey K.Editors}, year={2018},  collection={Cambridge Handbooks in Psychology}}

@article{carless2018development,
  title={The development of student feedback literacy: enabling uptake of feedback},
  author={Carless, David and Boud, David},
  journal={Assessment \& Evaluation in Higher Education},
  volume={43},
  number={8},
  year={2018},
  publisher={Taylor \& Francis}
}

@article{tay2022students,
  title={Students’ engagement across a typology of teacher feedback practices},
  author={Tay, Hui Yong and Lam, Karen WL},
  journal={Educational Research for Policy and Practice},
  volume={21},
  number={3},
  year={2022},
  publisher={Springer}
}

@article{fredricks2004school,
  title={School engagement: Potential of the concept, state of the evidence},
  author={Fredricks, Jennifer A and Blumenfeld, Phyllis C and Paris, Alison H},
  journal={Review of educational research},
  volume={74},
  number={1},
  year={2004},
  publisher={Sage Publications Sage CA: Thousand Oaks, CA}
}

@article{wong2024student,
  title={Student engagement and its association with academic achievement and subjective well-being: A systematic review and meta-analysis.},
  author={Wong, Zi Yang and Liem, Gregory Arief D and Chan, Melvin and Datu, Jesus Alfonso D},
  journal={Journal of Educational Psychology},
  volume={116},
  number={1},
  year={2024},
  publisher={American Psychological Association}
}

@article{wong2022student,
  title={Student engagement: Current state of the construct, conceptual refinement, and future research directions},
  author={Wong, Zi Yang and Liem, Gregory Arief D},
  journal={Educational Psychology Review},
  volume={34},
  number={1},
  year={2022},
  publisher={Springer}
}

@article{han2019learner,
  title={Learner engagement with written feedback: A sociocognitive perspective},
  author={Han, Ye and Hyland, Fiona},
  journal={Feedback in second language writing: Contexts and issues},
  pages={247--264},
  year={2019},
  publisher={Cambridge University Press Cambridge}
}

@article{henrie2015measuring,
  title={Measuring student engagement in technology-mediated learning: A review},
  author={Henrie, Curtis R and Halverson, Lisa R and Graham, Charles R},
  journal={Computers \& Education},
  volume={90},
  year={2015},
  publisher={Elsevier}
}

@article{martin2023integrating,
  title={Integrating motivation and instruction: Towards a unified approach in educational psychology},
  author={Martin, Andrew J},
  journal={Educational Psychology Review},
  volume={35},
  number={2},
  year={2023},
  publisher={Springer}
}

@article{miller2015using,
  title={Using reading times and eye-movements to measure cognitive engagement},
  author={Miller, Brian W},
  journal={Educational psychologist},
  volume={50},
  number={1},
  year={2015},
  publisher={Taylor \& Francis}
}

@inproceedings{niousha2026misconception,
  title={Misconception-Aware {LLM} Programming Tutor: Lessons Learned from Student-Tutor Interactions},
  author={Niousha, Rose and Boatright Smith, Samantha and O'Neill, Abigail and Zamfirescu-Pereira, JD and DeNero, John and Norouzi, Narges},
  booktitle={Proceedings of the 57th ACM Technical Symposium on Computer Science Education V. 2},
  year={2026}
}

\end{document}